\documentclass[twocolumn,preprintnumbers, superscriptaddress, reprint, prl]{revtex4-1}
\usepackage{amsmath}
\usepackage{stmaryrd}
\usepackage{txfonts}
\usepackage{amssymb}
\usepackage{mathrsfs}
\usepackage{graphicx}
\usepackage{dcolumn}
\usepackage{bm}
\usepackage{epsfig}
\usepackage{color}

\usepackage[colorlinks=true, linkcolor=blue, citecolor=blue, urlcolor=blue]{hyperref}
\begin{document}

\title{Interplay of the Spin Density Wave and a Possible Fulde-Ferrell-Larkin-Ovchinnikov State in $\mathrm{CeCoIn_5}$ in Rotating Magnetic Field}

\author{Shi-Zeng Lin}
\email{szl@lanl.gov}
\affiliation{Theoretical Division, T-4 and CNLS, Los Alamos National Laboratory, Los Alamos, New Mexico 87545, USA}
\author{Duk Y. Kim}
\email{duk0@skku.edu}
\affiliation{MPA-CMMS, Los Alamos National Laboratory, Los Alamos, New Mexico 87545, USA}
\affiliation{Center for Integrated Nanostructure Physics, Institute for Basic Science, Suwon, 16419, Republic of Korea}
\author{Eric D. Bauer}
\affiliation{MPA-CMMS, Los Alamos National Laboratory, Los Alamos, New Mexico 87545, USA}
\author{Filip Ronning}
\affiliation{MPA-CMMS, Los Alamos National Laboratory, Los Alamos, New Mexico 87545, USA}
\author{J. D. Thompson}
\affiliation{MPA-CMMS, Los Alamos National Laboratory, Los Alamos, New Mexico 87545, USA}
\author{Roman Movshovich}
\email{roman@lanl.gov}
\affiliation{MPA-CMMS, Los Alamos National Laboratory, Los Alamos, New Mexico 87545, USA}

\begin{abstract}

The $d$-wave superconductor $\mathrm{CeCoIn_5}$ has been proposed as a strong candidate for supporting the Fulde-Ferrell-Larkin-Ovchinnikov (FFLO) state near the low-temperature boundary of its upper critical field. Neutron diffraction, however, finds spin-density-wave (SDW) order in this part of the phase diagram for field in the $a$-$b$ plane, and evidence for the SDW disappears as the applied field is rotated toward the tetragonal $c$ axis. It is important to understand the interplay between the SDW and a possible FFLO state in  $\mathrm{CeCoIn_5}$, as the mere  existence of an SDW does not necessarily exclude an FFLO state.  Here, based on a model constructed on the basis of available experiments, we show that an FFLO state competes with an  SDW phase. The SDW state in $\mathrm{CeCoIn_5}$ is stabilized when the field is directed close to the $a$-$b$ plane. When the field is rotated toward the $c$ axis, the FFLO state emerges, and the SDW phase disappears. In the FFLO state, the nodal planes  with extra quasiparticles (where the superconducting order parameter is zero) are perpendicular to the field, and in the SDW phase, the quasiparticle density of states is reduced. We test this model prediction by measuring heat transported by normal quasiparticles in the superconducting state. As a function of field, we observe a reduction of thermal conductivity for field close to the $a$-$b$ plane and an enhancement of thermal conductivity when field is close to the $c$ axis, consistent with theoretical  expectations. Our modeling and experiments, therefore, indicate the existence of the FFLO state when field is parallel to the $c$ axis.

\end{abstract}

\date{\today}
\maketitle

\noindent
{\it Introduction.---} A key property of a superconductor is how it responds to an external magnetic field. Generally, magnetic field suppresses superconductivity. For superconductors with singlet Cooper pairing, magnetic field destroys superconductivity in two ways. First, the Lorentz force that an external magnetic field exerts on the two electrons of a Cooper pair have opposite directions. These forces tear a Cooper pair apart, thereby suppressing superconductivity via orbital limiting. Second, a magnetic field tends to polarize electron spins via Zeeman coupling, reducing the electrons' energy and leading to an enhanced Pauli susceptibility. Under certain circumstances, superconductivity can be destroyed by this mechanism of Pauli limiting even when Cooper pairs form with electrons of opposite spin alignment. While the maximal magnetic field that most known superconductors can sustain is defined by orbital limiting, some superconductors have been identified whose upper critical field is determined largely by Pauli limiting. These superconductors also can stabilize a spatially modulated superconducting state, known as the Fulde-Ferrel-Larkin-Ovchinnikov (FFLO) state, before superconductivity is suppressed entirely \cite{FFLOFF1964,FFLOLO1965}. In the FFLO state, Cooper pairs acquire a nonzero momentum, and the superconducting order parameter vanishes locally in space along nodal planes. This results in an excess of quasiparticles, which can significantly modify physical properties of the superconductors. Theoretically, it is clear that the formation of an FFLO state is unavoidable when a clean Pauli limited superconductor is subjected to a strong magnetic field. However, the experimental detection of an FFLO state remains a challenge despite some encouraging experimental evidence \cite{PhysRevLett.91.187004,FFLONMR2006,PhysRevLett.107.087002,PhysRevLett.109.027003,mayaffre_evidence_2014,PhysRevLett.116.067003,PhysRevLett.121.157004,PhysRevLett.119.217002,PhysRevB.85.174530,PhysRevB.97.144505,PhysRevLett.118.267001}, mainly because the superconducting order parameter cannot be measured directly. Often, competing effects render the interpretation of the experimental data difficult. 

The discovery of the heavy-fermion superconductor $\mathrm{CeCoIn_5}$ with tetragonal crystal structure and a transition temperature $T_c=2.3$ K has provided an exciting playground to search for the FFLO state. High quality $\mathrm{CeCoIn_5}$ with a large electron mean free path $l\approx \ 10 \xi$ can be achieved, where $\xi$ is the superconducting coherence length. Various measurements have revealed that the upper critical field in $\mathrm{CeCoIn_5}$ is mainly limited by the Pauli mechanism. Initial experimental measurements have shown the existence of a new phase \emph{inside} the superconducting phase for field both along the crystal $c$ and $a$ axis, see Ref. \onlinecite{matsuda_fuldeferrelllarkinovchinnikov_2007} for a review. For instance, a double-peak structure has been observed in NMR spectra in the high-field and low-temperature corner of the superconducting phase diagram \cite{FFLONMR2006}. In the case of $\mathbf{H}\parallel a$, neutron-scattering measurements have identified this new phase as a spin-density-wave (SDW) state that coexists with the superconducting state. In the SDW, magnetic moments with magnitude $0.1\ \mu_B$  ($\mu_B$ is the Bohr magneton) are aligned along the crystalline $c$ axis due to crystal field effects. A similar SDW is observed in the slightly Hg-doped $\mathrm{CeCoIn_5}$ \cite{PhysRevLett.121.037003}. This SDW is induced by magnetic field and disappears when superconductivity is destroyed by field. The SDW phase is suppressed (eventually entirely) when the field is rotated away from the $a$-$b$ plane \cite{PhysRevLett.105.187001}. The double peak structure in NMR spectra for $\mathbf{H}\parallel a$ can be explained in terms of the SDW phase. However the state responsible for similar NMR features for $\mathbf{H}\parallel c$ requires further study.

The appearance of an SDW only inside the superconducting phase comes as a surprise, because it is generally believed that SDW competes with superconductivity for the density of states at the Fermi surface. Several scenarios have been put forward to account for the stabilization of the SDW. These scenarios highlight the importance of the vortex lattice \cite{PhysRevB.83.140503}, Pauli pair breaking \cite{PhysRevB.82.060510,PhysRevB.83.224518}, the FFLO state \cite{yanase_antiferromagnetic_2009,PhysRevB.95.224513}, a superconducting pairing density wave \cite{PhysRevLett.102.207004, miyake_theory_2008,PhysRevLett.104.216403}, and Fermi surface nesting improved by the magnetic field \cite{PhysRevLett.107.096401,PhysRevB.86.174517,PhysRevB.92.224510}. It was shown in Ref. \onlinecite{PhysRevB.84.052508} that an in-plane magnetic field enhances the transverse magnetic susceptibility in Pauli-limited $d$-wave superconductors, which can result in a divergence of the dressed susceptibility. As a consequence, the magnetic fluctuations condense and static SDW order sets in. This magnon condensation picture is favored by inelastic neutron-scattering measurements \cite{PhysRevLett.100.087001,PhysRevLett.115.037001,PhysRevLett.109.167207}.

The emergence of an SDW does not rule out the existence of an FFLO state. In the case when the SDW competes with the FFLO state, one can suppress the SDW phase by rotating the field out of the $a$-$b$ plane (as is the case in $\mathrm{CeCoIn_5}$), which in turn can favor the FFLO state. This requires understanding the interplay between the SDW and the FFLO states, with a proper model that is relevant for $\mathrm{CeCoIn_5}$. In this Letter, we study the relation between SDW and FFLO states using a theoretical model described below.  We find that the SDW competes with the FFLO phase. When $\mathbf{H}\parallel a$, an SDW phase emerges inside the superconducting phase due to magnon condensation triggered by the magnetic field. When $\mathbf{H}$ is rotated towards the $c$ axis, the SDW phase is disfavored and the FFLO state appears. Because of the coupling between superconductivity and magnetism, the SDW order induces weak modulation in the superconducting order parameter and vice versa. Guided by the theory, we performed thermal-conductivity measurements in $\mathrm{CeCoIn_5}$ for a magnetic field applied within the $a$-$c$ plane as a function of angle away from the $c$ axis. The data are consistent with expectations from the model. The synergy between modeling and experiment suggests the presence of an FFLO phase in $\mathrm{CeCoIn_5}$ when  $\mathbf{H}\parallel c$.

\begin{figure}[t]
\psfig{figure=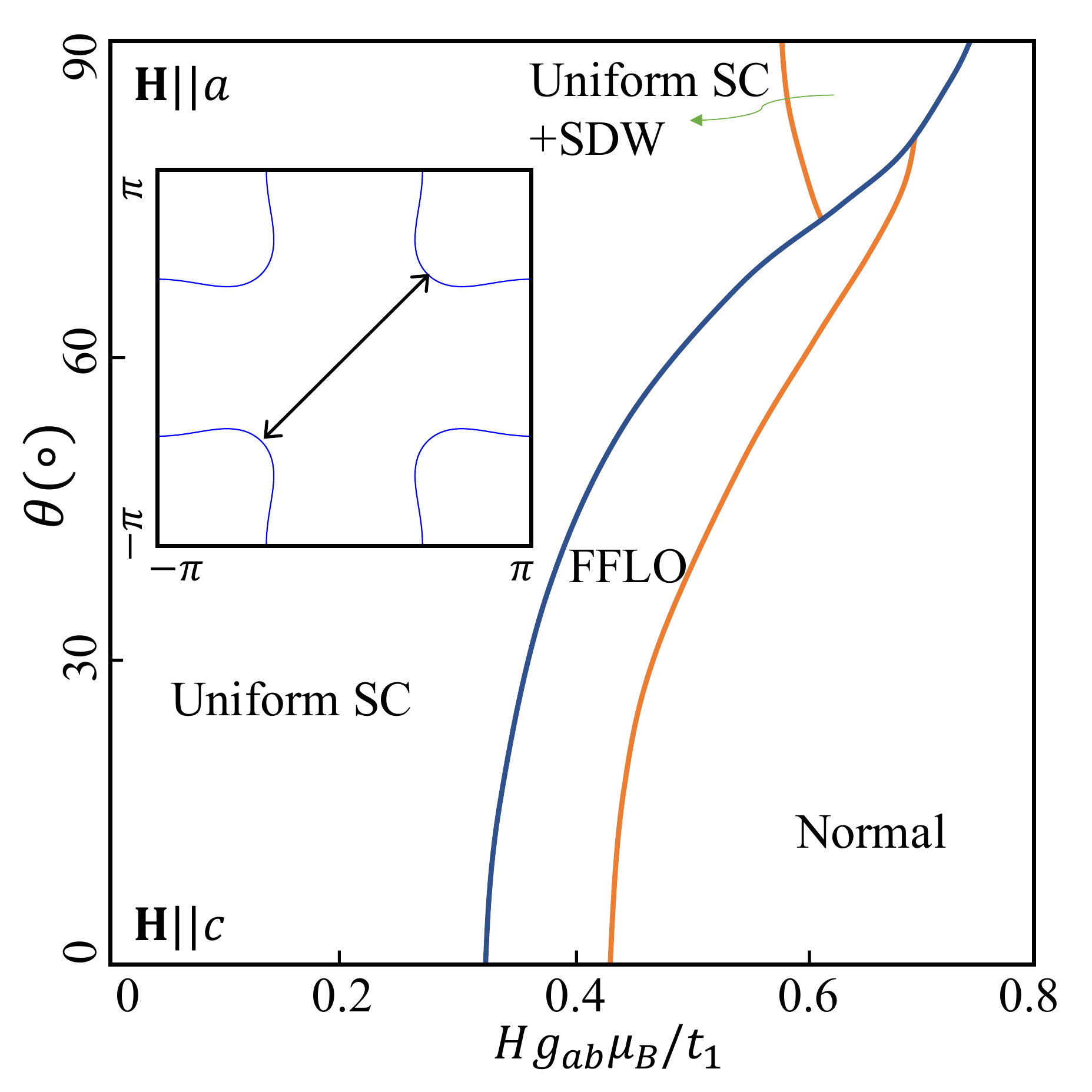,width=\columnwidth}
\caption{Theoretical phase diagram at $T=0$ when field is rotated in the $a$-$c$ plane. The phase diagram is constructed based on the FFLO and SDW order parameters. The orange (blue) line denotes a second (first) order phase transition. SC -- the superconducting state. The inset is the Fermi surface of the model in Eq. \eqref{eq1}. The arrow denotes the nesting wave vector.
} \label{f1}
\end{figure}

\begin{figure}[b]
\psfig{figure=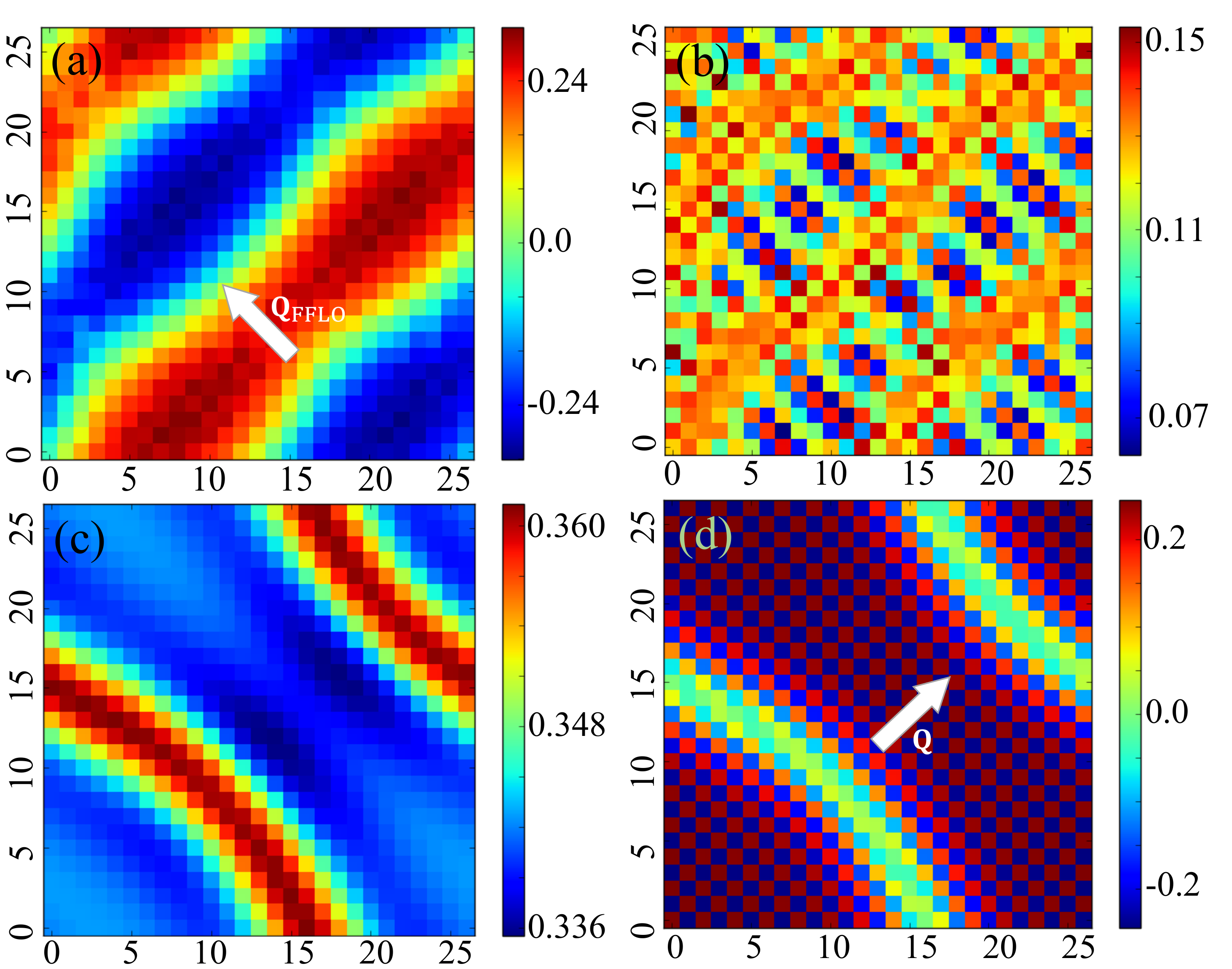,width=\columnwidth}
\caption{Profiles of superconducting and SDW order parameter $\Delta_d$ (a, c)  and $M_i=n_{i\uparrow}-n_{i\downarrow}$ (b, d). The upper (lower) two panels correspond to the FFLO (SDW) state at $g_{ab}\mu_B H=0.4t_1$ and $\mathbf{H}\parallel c$ ($g_{ab}\mu_B H=0.6 t_1$ and $\mathbf{H}\parallel a$). The horizontal and vertical axes are the tight-binding lattice sites. $\Delta_d$ is in unit of $t_1$. The arrows in (a) and (d) are the direction of the wave vector of the FFLO and SDW.
} \label{f2}
\end{figure}

\noindent
{\it Model.---}
We construct a model Hamiltonian based on the following experimentally established facts. (1) Various experiments have shown that $\mathrm{CeCoIn_5}$ is close to an SDW instability \cite{PhysRevLett.91.257001,PhysRevLett.91.246405,PhysRevB.73.064519,PhysRevLett.108.056401,doi:10.1143/JPSJ.72.2308,PhysRevLett.100.087001,PhysRevLett.115.037001,PhysRevLett.109.167207}, which is consistent with the fact that a field of order of $11$ T is sufficient to trigger an SDW instability. (2) The SDW is formed by gapping quasiparticles in  nodes of the superconducting $d_{x^2-y^2}$ order parameter. The SDW ordering wave vector is $\mathbf{Q}=(0.44,\ \pm 0.44,\ 0.5)$. (3) The SDW moments have strong Ising anisotropy and lie along the $c$ axis. (4) The Fermi surface relevant for superconductivity is a warped cylinder. Based on these facts, we can construct the following mean-field Hamiltonian in two dimensions:        
\begin{align}\label{eq1}
\begin{split}
{\cal H} = \sum_{i, j,\sigma}t_{ij}c_{i\sigma}^{\dagger}c_{j\sigma}-\mu\sum_{i,\sigma}c_{i\sigma}^{\dagger}c_{i\sigma}+\sum_{\langle i, j\rangle }(\Delta_{ij}c_{i\uparrow}^{\dagger}c_{j\downarrow}^{\dagger}+\Delta_{ij}^*c_{j\downarrow}c_{i\uparrow})\\
-\sum_{i}\left(h_{i} s_{z, i}-g_{ab}\mu_B H_x  s_{x, i}-g_{c}\mu_B H_z  s_{z, i}\right).
\end{split}    
\end{align}
where $t_{ij}$ describes electron hopping on a square lattice with the dispersion $\epsilon(\mathbf{k})=2 t_1 [\cos(k_x)+\cos(k_y)]+4 t_2\cos(k_x)\cos(k_y)+2t_3[\cos(2k_x)+\cos(2k_y)]-\mu$. Here $t_1$, $t_2=-0.5 t_1$ and $t_3=-0.4 t_1$ are the nearest neighbor (NN), the second NN (along the diagonal), and the third NN (along the bond) hopping amplitudes, respectively, and their strengths have been estimated by using density functional theory \cite{tanaka_theory_2006}. Here $\mu_B$ in Eq. \eqref{eq1} is the Bohr magneton. The electron density for up and down spins is $n_{i\uparrow}=\sum_l|u_{i\uparrow,l}|^2 f(E_l)$ and $n_{i\downarrow}=\sum_l |v_{i\downarrow,l}|^2 f(-E_l)$, where $f(E_l)$ is the Fermi distribution function and $u_{i\uparrow,l}$, $v_{i\downarrow,l}$, $E_l$ are the $l$th eigenvector and eigenenergy of the Bogoliubov-de Gennes equation associated with Eq. \eqref{eq1} \cite{JXZhuBook}. We fix the electronic density at $\langle n\rangle=0.72$ by tuning the chemical potential $\mu$. 

The self-consistent equation for the pairing potential is $\Delta_{ij}$ and for molecular field $h_i$ is $\Delta_{ij}=\frac{V}{4}\sum_l\left(u_{i\uparrow,l}v_{j\downarrow,l}^*+u_{j\uparrow,l}v_{i\downarrow,l}^*\right)\tanh(E_l/2T)$ and $h_i=-\sum_{j} J_{ij}(n_{j\uparrow}-n_{j\downarrow})$. The spin of conduction electrons is $\mathbf{s}_{i}=\sum_{\alpha'\beta'}c_{i\alpha'}^{\dagger}\mathbf{\sigma}_{\alpha'\beta'}c_{i\beta'}$. There is a strong anisotropy in the electron $g$ factor \cite{PhysRevB.71.104528}, and we take $g_c/g_{ab}=3$. To stabilize SDW order at $\mathbf{Q}$, we use $J_{ij}=J_1$ (NN antiferromagnetic interaction) and the third NN (competing) interaction  $J_3=-J_1/4\cos(2\pi Q_x)$. To ensure $d$-wave pairing symmetry, we choose the NN pairing potential $\Delta_{ij}$ with $V=4.5 t_1$ in the calculations. The $d$-wave order parameter is given by $\Delta_d=(\Delta_{i,i+\hat{x}}+\Delta_{i, i-\hat{x}}-\Delta_{i,i+\hat{y}}-\Delta_{i, i-\hat{y}})/4$, where $\hat{x}$ and $\hat{y}$ are the unit vectors in the $x$ and $y$ direction, respectively \cite{PhysRevB.50.13883}. Similar models were introduced to describe the emergence of an SDW state in Pauli-limited superconductors in high field \cite{PhysRevLett.107.096401,PhysRevB.86.174517}, as well as the phase diagram for Nd-doped $\mathrm{CeCoIn_5}$ \cite{PhysRevB.92.224510,SZLinZhu2017}.

This model describes the competition between an SDW and $d$-wave superconductivity, which depends on values of $V$ and $J_1$ \cite{RevModPhys.62.113}. When $J_1/t_1>3.6$, SDW order develops and coexists with superconducting order \cite{SZLinZhu2017}. To model $\mathrm{CeCoIn_5}$, the system is tuned to an SDW instability by setting $J_1/t_1=3.2$. The phase diagram at zero temperature $T=0$, when field is rotated in the $a$-$c$ plane, is displayed in Fig. \ref{f1}. For $\mathbf{H}\parallel a$, the model correctly describes the development of the SDW phase inside the superconducting state at high field. When field is canted towards the $c$ axis, the SDW state is suppressed, and the FFLO state appears \footnote{In the calculation with finite system size, there are several FFLO states with different wave vector $Q$ separated by the normal state at higher fields because of the commensuration of $Q$ of the FFLO state with the system size. See T. Kim, C.-C. Chien and S. -Z. Lin, Phys. Rev. B $\mathbf{99}$, 054509 (2019).}. The transition from the uniform superconducting state to the FFLO (SDW) state is of first (second) order, while the transition from the FFLO state (SDW) to the normal state is of second (first) order. The spatial distributions of $\Delta_d$ and SDW orders are shown in Fig. \ref{f2}. In the FFLO state, the modulation of $\Delta_d$ induces modulation of magnetization, see Figs. \ref{f2}(a) and \ref{f2}(b). The relatively weak magnetization modulation follows the modulation of $\Delta_d$, with magnetization maximal in the nodes (zeros) of the FFLO phase. There is an additional modulation of the magnetization in the region where the amplitude of $\Delta_d$ is maximal. While in the SDW phase, modulation of SDW order generates weak modulation in $\Delta_d$, see Figs. \ref{f2}(c) and \ref{f2}(d). Because of the competition between SDW and superconductivity, the maxima in $\Delta_d$ corresponds to the minima in the SDW order, and vice versa.

In our 2D calculations, the wave vector of the FFLO, $\mathbf{Q}_{\mathrm{FFLO}}$, is confined to the plane. For an $s$-wave superconductor with a vortex lattice,  $\mathbf{Q}_{\mathrm{FFLO}}$ is parallel to the vortex lines and $\mathbf{H}$ \cite{PhysRevLett.16.996,PhysRevLett.95.117003}. For $d$ wave, one would also expect that $\mathbf{Q}_{\mathrm{FFLO}}\parallel\mathbf{H}$, otherwise there would be modulation of the superconducting order parameter with two different periods in the $a$-$b$ plane originating from the FFLO and vortex lattice. This is energetically disfavored both by Pauli and orbital pair breaking effects. This picture is supported by numerical calculation based on a quasiclassical theory for a single vortex in a $d$-wave superconductor \cite{PhysRevB.76.014503}. For  $\mathbf{Q}_{\mathrm{FFLO}}\parallel\mathbf{H}$, the thermal conductivity will be enhanced when thermal current $\mathbf{J}\perp \mathbf{H}$ due to the excess of quasiparticles around nodal planes in the FFLO state. In contrast, some of nodal quasiparticles of the $d$-wave superconductivity are gapped out in the SDW state and one expects a reduction of the thermal conductivity. These expectations are borne out by the thermal-conductivity measurements.

\begin{figure}[t]
\psfig{figure=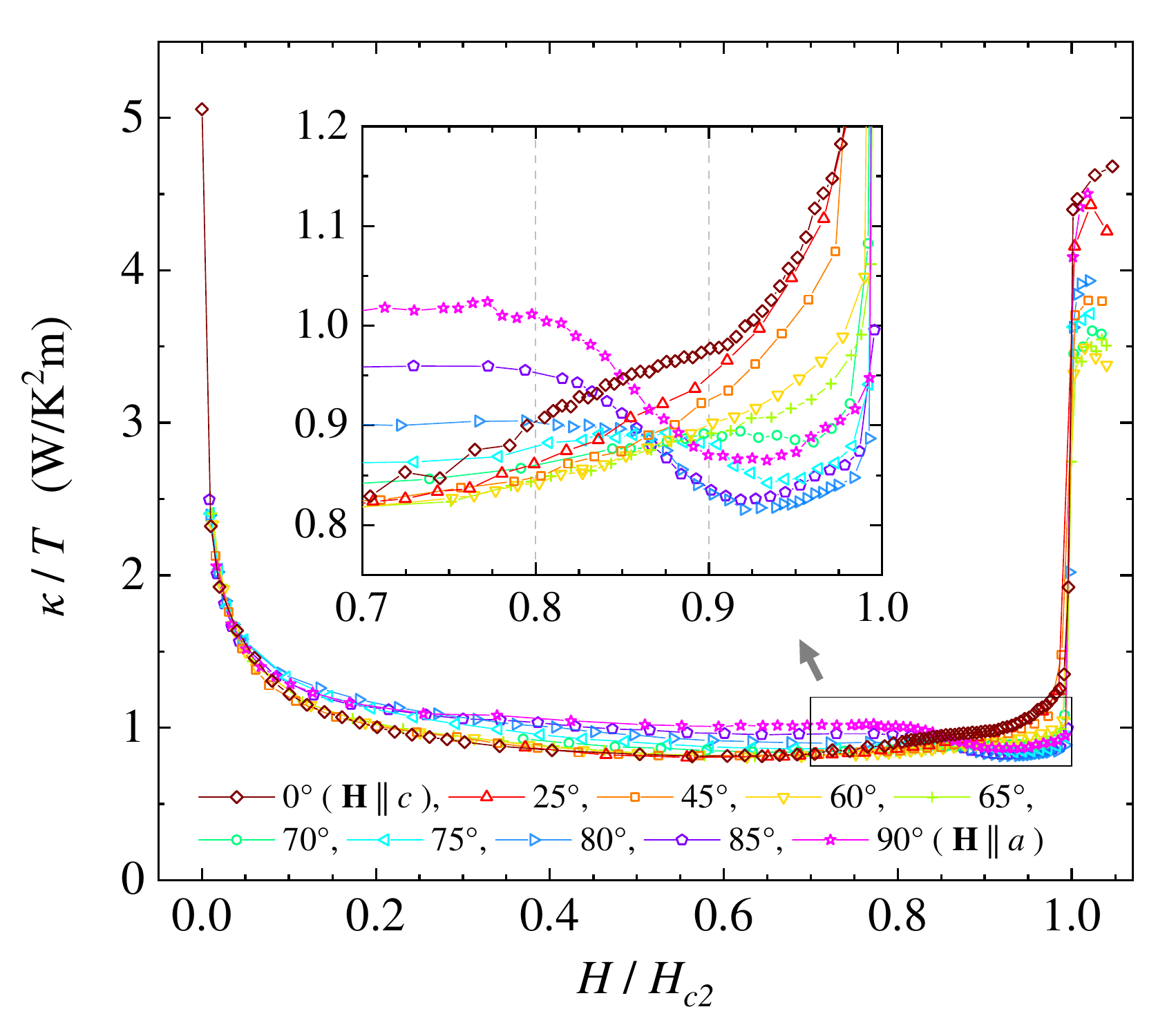,width=\columnwidth}
\caption{Thermal conductivity with $\mathbf{J}\parallel b$ as a function of the magnetic-field strength at different field directions. The angle in the label represents the difference between the field direction and the crystalographic $c$ axis. The inset shows the expanded view of the data near $H_{c2}$. The measurement temperature $T$ is 0.09 K.
} \label{f3}
\end{figure}

\begin{figure}[b]
\psfig{figure=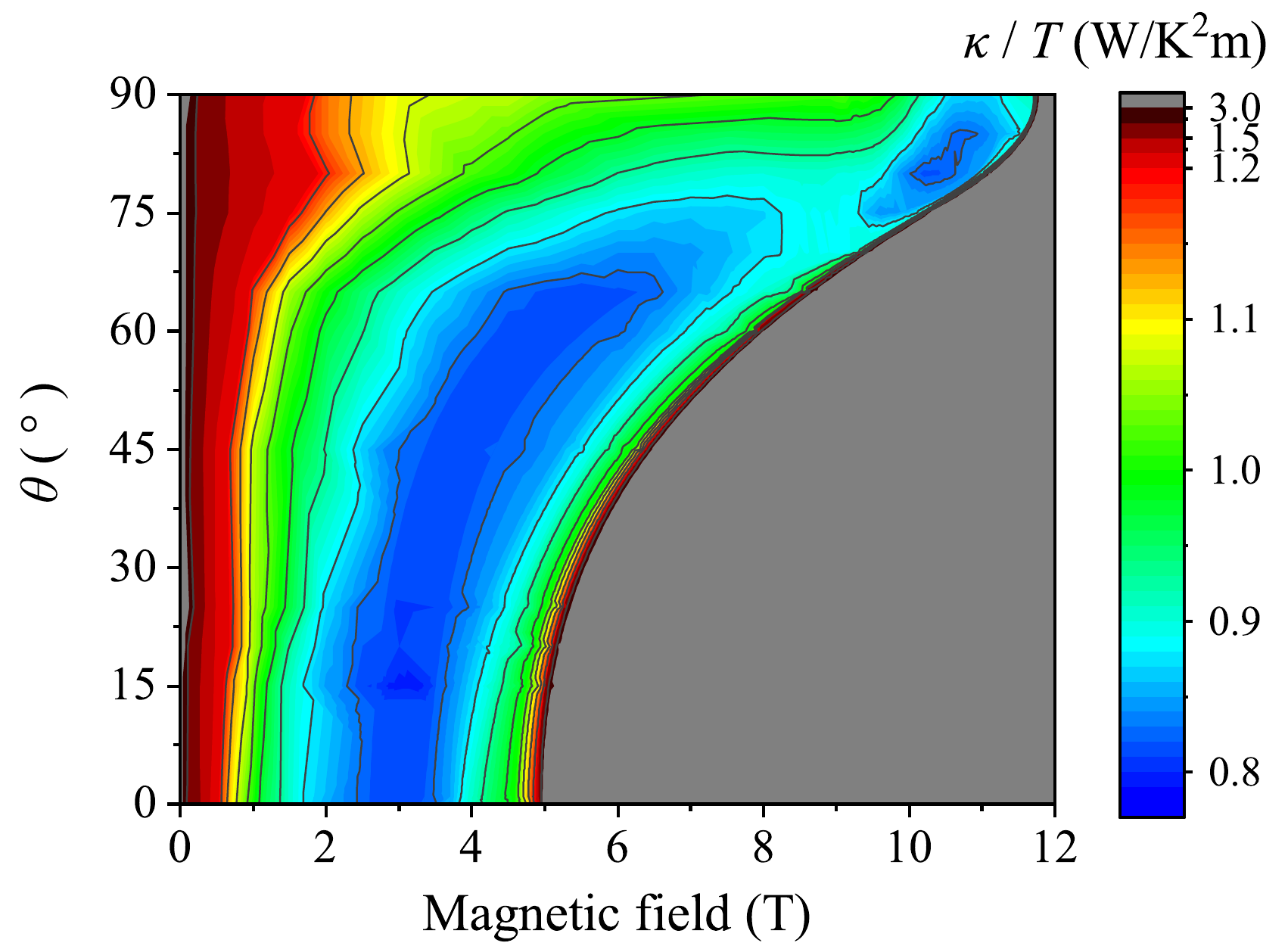,width=\columnwidth}
\caption{Contour plot of thermal conductivity. $T$=0.09 K. The black curve separating colored and gray areas is $H_{c2}(\theta)$. The contour plot was constructed from field-scan data at fixed angles of $\theta$ = 0$^{\circ}$, 15$^{\circ}$, 20$^{\circ}$, 25$^{\circ}$, 45$^{\circ}$, 60$^{\circ}$, 65$^{\circ}$, 70$^{\circ}$, 75$^{\circ}$, 80$^{\circ}$, 85$^{\circ}$, and 90$^{\circ}$. Wide blue region for low angles represents a slowly varying background (as seen in Fig. \ref{f3}) between roughly 1.5 and 4 T for $\mathbf{H}\parallel c$. Between this flat region and a critical field $H_{c2}$, the thermal conductivity rises sharply (green to red strip) indicating an FFLO phase.
} \label{f4}
\end{figure}

\noindent
{\it Thermal conductivity measurements.---}
We performed thermal-conductivity measurements on $\mathrm{CeCoIn_5}$ with magnetic field rotating from $\mathbf{H}\parallel c$ to $\mathbf{H}\parallel a$ at $T=0.09$ K. To minimize the effect of the vortex lattice on the thermal conductivity, we chose  configuration where the vortex lattice is always perpendicular to the thermal current during field rotation, i.e. $\mathbf{J}\parallel b$ and field in the $a$-$c$ plane. Single crystal $\mathrm{CeCoIn_5}$  ($0.2\times2.5\times0.05~\mathrm{mm}^3$) was grown from an excess indium flux, and the thermal conductivity was measured by the standard steady-state method. The results are shown in Fig. \ref{f3}. $\kappa$ initially decreases with field due to vortex scattering. This behavior indicates a high quality of the crystal because the electronic mean free path is mainly limited by vortex scattering even with very large intervortex distances at low fields. For $\mathbf{H}$ close to the $a$-$b$ plane ($\theta\sim 90^{\circ}$), the thermal conductivity $\kappa$ drops with increasing field before reaching $H_{c2}$. This drop originates from the development of the SDW and concomitant pair-density-wave (PDW) orders, which gap out some of $d$-wave nodal quasiparticles \cite{Kim2016} \footnote{The PDW is an induced order, and is of secondary importance in current discussions. The FFLO state is a dominant order when $\mathbf{H}\parallel c$.}. As field is rotated toward the $c$ axis, the drop decreases, signaling the suppression of SDW order. The threshold angle where this sharp drop disappears is approximately $70^\circ$, which is consistent with previous measurements \cite{PhysRevLett.105.187001,PhysRevB.71.020503}. For $\mathbf{H}$ close to the $c$ axis, a significant increase of $\kappa$ is observed as $H$ approaches $H_{c2}$ from below. We ascribe this enhancement to a signature of the FFLO state, in accord with the model calculation. Such enhancement of thermal conductivity below $H_{c2}$ is not expected for a Pauli-limited superconductor with a first-order superconducting transition. Instead, thermal conductivity below $H_{c2}$ should be rather field independent up to $H_{c2}$, and display a sharp step at $H_{c2}$. A contour plot of $\kappa$ in the complete field and angle range is depicted in Fig. \ref{f4}.  The behavior of $\kappa$ near $H_{c2}$ is consistent with the expectation from our model.

\noindent
{\it Discussions and summary.---}
We stress the importance of Ising-like magnetic anisotropy in the model. The presence of this anisotropy naturally explains why the SDW phase only occurs for field close to the $a$-$b$ plane within the magnon condensation picture \cite{PhysRevB.84.052508}. It also leads to the competition between the SDW and FFLO states, as shown in Fig. \ref{f1}. Our approach and results are different from a model with isotropic magnetic fluctuations \cite{yanase_antiferromagnetic_2009}, where it was proposed that the SDW phase is stabilized by Andreev bound states localized around FFLO nodal planes. We too observe an induced magnetization oscillation in the FFLO state (see Fig. \ref{f2}), but with a much weaker amplitude compared to that for $\mathbf{H}\parallel a$, where the full SDW develops. In addition, the picture that the SDW state for $\mathbf{H}\parallel a$ is induced by the FFLO state is not supported by recent neutron-scattering measurements, which reveal that the SDW state is induced by closing the magnon gap \cite{PhysRevLett.100.087001,PhysRevLett.115.037001,PhysRevLett.109.167207}. The different phenomenology for $\mathbf{H}||ab$ and $\mathbf{H}||c$ highlights the importance of spin anisotropy.

A number  of measurements reveal the presence of a quantum critical point around $H_{c2}=5$ T when $\mathbf{H}\parallel c$ \cite{PhysRevLett.91.246405,PhysRevLett.97.106606,PhysRevLett.117.016601}. Strong quantum fluctuations should suppress the thermal conductivity near $H_{c2}$. There must exist another mechanism which counters this suppression and leads to the enhancement of $\kappa$, as observed experimentally. Model calculations show that $\kappa$ increases with $H$ near $H_{c2}$ in orbital-limited superconductors \cite{PhysRevLett.83.4626}. The orbital-limited critical field $H_{c2}^\mathrm{orb}$ in $\mathrm{CeCoIn_5}$ is about three times larger than the Pauli limited critical field $H_{c2}^P$, and therefore, the experimentally measured superconducting critical field $H_{c2}$ as well. The effect of orbital limiting on thermal conductivity near $H_{c2}$, can be gleaned from calculations for the case of orbital limiting alone around a field of about ${1\over 3}H_{c2}$ \cite{PhysRevB.75.224501,PhysRevB.75.224502}. The variation of thermal conductivity with field is very slow in this region, in contrast to the sharp increase in the thermal conductivity near $H_{c2}$ shown in Fig. \ref{f3}. Therefore, orbital limiting cannot be causing the observed increase in $\kappa$.  The vortex lattice undergoes a number of structural transitions as a function of field strength when $\mathbf{H}\parallel c$, including transition from square to rhombic to triangular between 3 T and $H_{c2} = 5$ T \cite{bianchi_superconducting_2008}. The vortex lattice transition is a gradual process. For instance, the apex angle of a unit cell for triangle vortex lattice is $60^\circ$ and that for a square lattice is $90^\circ$. The evolution from triangular to the square lattices corresponds to the continuous change of the apex angle from $60^\circ$ to $90^\circ$ \cite{PhysRevB.95.054511}. The smooth deformation of the vortex lattice, while the vortex lines are kept perpendicular to heat current, is therefore unlikely to give rise to a dramatic increases of $\kappa$. The Pauli pairing breaking effect, therefore, should be dominant in CeCoIn$_5$ at fields near $H_{c2}$.   

While our modeling and experiments suggest the existence of the FFLO state when the field is orientated close to the $c$ axis, two experimental observations deserve further attention. First, the region of enhanced $\kappa$ near $H_{c2}$ is wider than that identified as the FFLO region by NMR measurements \cite{FFLONMR2006}. However, a similar effect was observed for the field dependence of thermal conductivity in CeCoIn$_5$ at the SDW transition for field in the {\it a-b} plane \cite{Kim2016}. There, the onset of the reduction in thermal conductivity at very low temperature was observed to take place at 9 T, whereas specific heat and neutron scattering measurements firmly place the SDW transition at 10 T. This may be due to fluctuations of the order parameter. Second, we are not able to resolve hysteresis in $\kappa$ near $H_{c2}$ when the direction of the field sweep is reversed. Such hysteresis is expected from the first order phase transition between the uniform superconducting state and the FFLO state. The lack of hysteresis may be due to a weak first order nature of the transition. We remark that the model calculations show a second order phase transition at $H_{c2}$ while it is a first order transition in experiments. However, the nature of the phase transition at $H_{c2}$ depends on the dimensionality and microscopic details of the model \cite{buzdin_generalized_1997,PhysRevB.63.184521}, which may cause the discrepancy.  

To summarize, combined modeling and thermal-conductivity measurements suggest the existence of an FFLO state in CeCoIn$_5$ for field aligned along the $c$ axis. The FFLO state competes with an SDW phase, and their relative stability can be tuned by rotating field in the $a$-$c$ plane. Additional neutron scattering measurements with a scattering plane that includes the $c$ axis may be promising in resolving the modulated susceptibility caused by the FFLO state. We expect the FFLO modulation wave vector to lie along the $c$ axis when magnetic field is applied along the $c$ axis. We note that the anomalous drop in the vortex lattice form factor, observed by the neutron scattering for field close to $H_{c2}$, was ascribed to the formation of an FFLO state \cite{bianchi_superconducting_2008}.

\begin{acknowledgements}

The authors would like to thank Takanori Taniguchi and Daniel Mazzone for helpful discussions. Computer resources for numerical calculations were supported by the Institutional Computing Program at LANL. The theoretical work by S. Z. L. was carried out under the auspices of the U.S. DOE Contract No. DE-AC52-06NA25396 through the LDRD program. Sample synthesis was supported by the DOE BES ``Quantum Fluctuations in Narrow Band Systems'' project. The thermal conductivity measurements were supported by the LDRD program at LANL. D. Y. K. acknowledges the Young Scientist Fellowship from the Institute for Basic Science.

\end{acknowledgements}

\bibliography{CeCoIn5FFLOSDW}

\end{document}